\documentclass[acus]{JAC2003}

\usepackage{graphicx}
\usepackage{booktabs}
\usepackage{relsize}
\newcommand{\ev}{{\rm e}\kern-1.pt{\rm V}}
\newcommand{\gev}{{\rm Ge}\kern-1.pt{\rm V}}
\newcommand{\mev}{{\rm Me}\kern-1.pt{\rm V}}
\newcommand{\kev}{{\rm ke}\kern-1.pt{\rm V}}
\newcommand{\tev}{{\rm Te}\kern-1.pt{\rm V}}
\newcommand{\gevsq}{\mbox{$\mathrm{{\rm Ge}\kern-1.pt{\rm V}}^2$}}

\def\cesrta{{C{\smaller[2]ESR}TA}}
\def\lsim{\mathrel{\rlap{\lower4pt\hbox{\hskip1pt$\sim$}}
    \raise2pt\hbox{$<$}}} 
\def\gsim{\mathrel{\rlap{\lower4pt\hbox{\hskip1pt$\sim$}}
    \raise2pt\hbox{$>$}}} 

\begin{document}
\title{
Electron Cloud Buildup Characterization Using Shielded Pickup
Measurements and Custom Modeling Code at {\cesrta}
}
\author{J.A.~Crittenden and J.P.~Sikora\\ 
CLASSE\thanks{
Work supported by the US National Science Foundation 
(\mbox{PHY-0734867}, \mbox{PHY-1002467}, and \mbox{PHY-1068662}), US Department of Energy
(\mbox{DE-FC02-08ER41538}), and the Japan/US Cooperation Program
},
Cornell University, Ithaca, NY 14850, USA
}

\maketitle

\begin{abstract}
The Cornell Electron Storage Ring Test Accelerator
experimental program
includes investigations into electron cloud buildup, applying various 
mitigation techniques in custom vacuum chambers. Among these are two 
1.1-m-long sections located symmetrically in the east and west arc 
regions. These chambers are equipped with pickup detectors shielded 
against the direct beam-induced signal. They detect cloud electrons 
migrating through an 18-mm-diameter pattern of small holes in the top of 
the chamber. A digitizing oscilloscope is used to record the signals, 
providing time-resolved information on cloud development. Carbon-coated, 
TiN-coated and uncoated aluminum chambers have been tested. Electron 
and positron beams of 2.1, 4.0 and 5.3~{\gev} with a variety of bunch 
populations and spacings in steps of 4 and 14 ns have been used. 
Here we report on results from the ECLOUD modeling code which highlight 
the sensitivity of these measurements to the physical phenomena
determining cloud buildup such as the 
photoelectron production azimuthal and energy distributions, and the 
secondary yield parameters including the true secondary, re-diffused, 
and elastic yield values.
\end{abstract}

\section{Introduction}
\begin{figure}[htbp]
   \centering
   \includegraphics*[width=70mm]{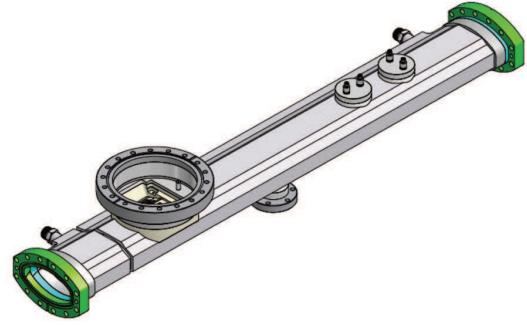}
   \caption{Custom vacuum chamber with RFA port and shielded pickup detectors.}
   \vspace{-1.1mm}
   \label{fig:vc}
\end{figure}

\begin{figure}[htbp]
   \centering
   \includegraphics*[width=78mm]{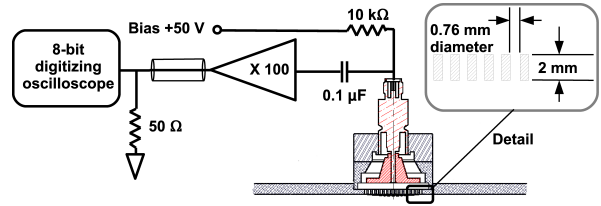}
   \caption{SPU detector design and readout. The 3:1 ratio of depth to diameter
of the port holes in the top of the beam-pipe effectively shields the BPM-style
collector electrode (button) from the direct beam signal. The 50-V positive
bias serves to prevent secondary electrons produced on the button from escaping.
The signals are typically
digitized with 8-bit accuracy in 0.1-ns steps over 100-ns, averaging over 8k triggers.
}
   \vspace{-1.1mm}
   \label{fig:spudesign}
\end{figure}

The Cornell Electron Storage Ring Test Accelerator ({\cesrta}) project~\cite{ref:icfa50}  
has been
exploiting the versatility of the 768-m-circumference CESR storage ring to obtain
measurements of low-emittance beams and electron cloud buildup for electron and
positron beams ranging from 1.8 to 5.3~GeV. The program 
includes the installation of custom vacuum chambers
with retarding-field-analyzer (RFA) ports and shielded pickup (SPU) detectors of the
type shown in Fig.~\ref{fig:vc}. 
The RFA port is shown on the left end,
and two circular SPU modules are shown on the right end of the
chamber, each with two ports. In one case the two ports are placed longitudinally,
and in the other case the two ports
are arranged transversely, providing laterally segmented sensitivity to the 
cloud electrons. Thus the centers of buttons are 0, and ${\pm 14}$~mm from the
horizontal center of the chamber.
The ports consist of 169 0.76-mm-diameter holes arranged in concentric circles
up to a maximum diameter of 18~mm. The top of the vacuum chamber has
been machined such that the holes are aligned vertically. 
The transparency factor for vertical trajectories is ~27\%. The
approximate 3:1 depth-to-diameter factor is chosen to effectively shield
the detectors from the signal induced directly by the beam, as shown in 
Fig.~\ref{fig:spudesign}.

Time-resolved measurements provide time structure information
on electron cloud (EC) development, in contrast to the time-integrated 
RFA measurements~\cite{ref:ecloud10jrc}.
However, they have relatively primitive energy selection, since they have no retarding grid.
Also, the position segmentation is more coarse, the charge-collecting electrodes
being of diameter 18 mm. Data has been recorded with biases of 0 and
$\pm 50$~V relative to the vacuum chamber. 
The studies described here address exclusively the data
with bias $+50$~V in order to avoid contributions to the signal from secondary
electrons escaping the pickup. Such secondaries generally carry kinetic energy insufficient
to escape a 50~V bias. This choice of bias obviously provides sensitivity to
cloud electrons which enter the port holes with low kinetic energy.
The front-end readout electronics comprise two Mini-Circuits ZFL-500 broadband amplifiers
with $50~{\Omega}$ input impedance for a total gain of 40~dB.
Digitized oscilloscope traces are recorded with 0.1~ns step size to 8-bit accuracy 
with auto-scaling, averaging over 8k triggers.

\begin{figure}[htbp]
   \centering
   \includegraphics*[width=78mm]{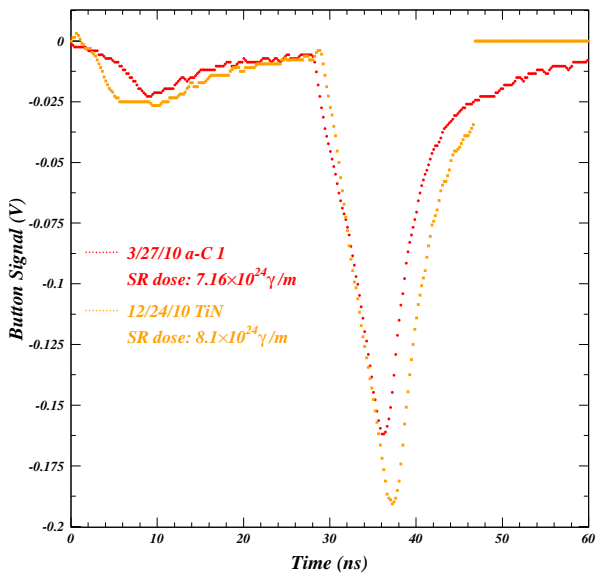}
   \caption{Examples of SPU signals produced by two bunches of \mbox{$8{\times}10^{10}$} 5.3~GeV positrons 
spaced 28~ns apart. Signals recorded in March and December of 2010 for 
amorphous-carbon-coated and TiN-coated aluminum
vacuum chambers at the same position in CESR ring are compared. 
The  chambers have 
each been well-conditioned by a synchrotron radiation dose of about \mbox{$8{\times}10^{24} {\gamma}/m$}. 
}
   \vspace{-1.1mm}
   \label{fig:ex1}
\end{figure}

\begin{figure}[htbp]
   \centering
   \includegraphics*[width=78mm]{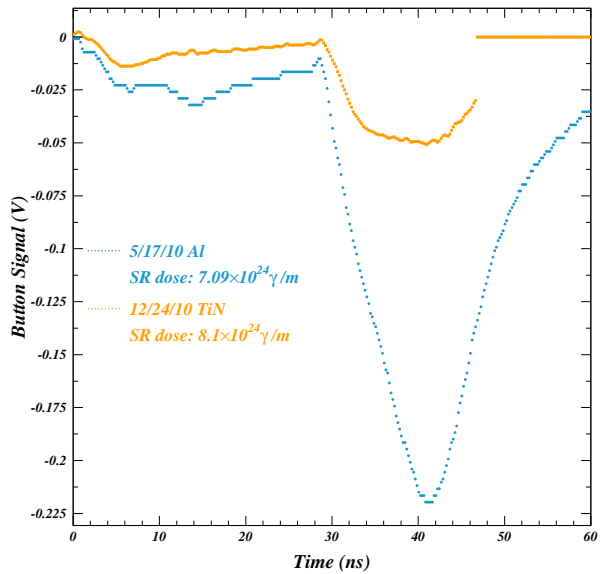}
   \caption{Comparison of signals recorded in May and December 2010 in 
an uncoated aluminum chamber and in a TiN-coated chamber, whereby the populations of the two 
positron bunches are \mbox{$4.8{\times}10^{10}$}.
The larger signal in the uncoated chamber indicates much greater cloud buildup. Note also that 
the reduced bunch population relative to Fig.~\ref{fig:ex1} results in a much smaller signal
in the TiN-coated chamber as well as a different time dependence.
}
   \vspace{-1.1mm}
   \label{fig:ex2}
\end{figure}

Figures~\ref{fig:ex1} and~\ref{fig:ex2} 
show examples of digitized SPU signals produced by two positron bunches spaced 28~ns apart.
Anticipating the following discussion of the interpretation of these signals, we note that
the time characteristics of these signals carry much detailed information on EC development. 
The leading bunch seeds the cloud and produces photoelectrons which drift into the SPU
detector. The leading signal is produced by the photoelectrons produced on the bottom of the vacuum 
chamber, since they are the first to arrive at the top of the chamber, accelerated by the positron 
bunch directly at the detector. The second signal peak is larger, since it carries a contribution 
from the cloud present below the beam at the time of arrival of the second bunch. Since these cloud 
electrons have been produced by wall interactions during the preceding 28~ns, the size and shape of 
this second signal peak depends directly on the secondary yield characteristics of the vacuum 
chamber surface. The variety of signal shapes and magnitudes for the two-bunch measurements 
shown in Figs.~\ref{fig:ex1} and ~\ref{fig:ex2} make clear that detailed information on all aspects of EC
buildup in different mitigation environments can be obtained. Below we describe the numerical 
modeling which instructs our interpretation of these measurements.  

\section{Numerical Modeling of EC Buildup}
The EC buildup modeling code ECLOUD~\cite{ref:icfazim} has been under active
development for the purposes of {\cesrta} since 2008. Developed at CERN in the 1990s, 
it has seen widespread application for EC phenomena observed at the CERN LHC, SPS, 
and PS, as well as at KEK and RHIC. It has been extensively 
benchmarked~\cite{ref:pac2009jrcbench} against the 2D buildup code 
POSINST~\cite{ref:furmanpivi} and has successfully described the 
{\cesrta} measurements of EC-induced coherent tune 
shifts~\cite{ref:ecloud12gfd,ref:ipac11dlk,ref:ecloud10dlk,ref:ipac10jac,ref:pac09jac}. 
ECLOUD includes simulation algorithms 
for photoelectron generation,
for time-sliced macroparticle tracking in the 2D electrostatic fields sourced by the beam and 
the cloud, and 3D tracking in a variety of ambient magnetic fields, as well as for a detailed
model of the interactions of cloud electrons with the vacuum chamber surface producing secondary
electrons. A variety of options have been implemented to model the {\cesrta} measurements.
The azimuthal distribution of photoelectron generation sites for the modeling results
presented in this paper are provided by the recently developed photon reflection and
tracking modeling code Synrad3D~\cite{ref:ecloud10gfd}, which calculates photon rates and
absorption sites throughout the CESR lattice using a detailed model 
of the vacuum chamber around the entire ring and a given set of vacuum chamber surface roughness
parameters. The photoelectron generation portion of the ECLOUD code has been generalized 
to allow admixtures of various power-law photoelectron energy distributions in addition
to the Gaussian functions originally provided. The model for the generation of secondary
electrons has been generalized to allow the same set of parameters used in the POSINST code.
Response functions for the SPU detector have been implemented. As a function
of incident angle and energy, a fraction of
a macroparticle charge hitting the wall in the region of the detector on the top of
the beam-pipe contributes to the modeled signal. The remaining charge can generate secondary electrons. 
A contribution
to the signal from secondaries generated on the walls of the 0.76-mm-diameter SPU holes is also
calculated. The modeled signal in each time slice thus carries a statistical error associated
with the number of contributing macroparticles. Typically \mbox{$2{\times}10^6$} macroparticles are
generated during the passage of each bunch, each macroparticle carrying thousands
of electron charges. The development of the cloud is calculated in
100 time slices during the passage of the 1-cm long bunch, and in 2000 time slices between
bunch passages. 

\section{Photoelectron Model}

\begin{figure}[b]
   \centering
   \includegraphics*[width=78mm]{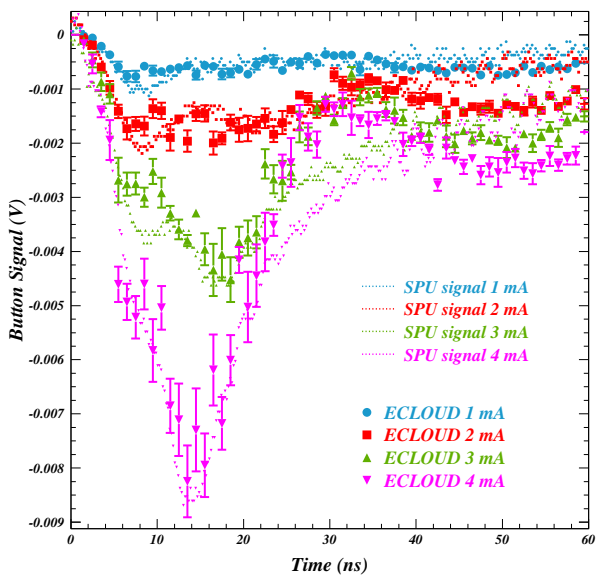}
   \caption{Single bunch signal dependence on positron bunch population. The small symbols
show the digitized SPU signal. The larger symbols with error bars show the ECLOUD-modeled signal
and its associated statistical uncertainty arising from counting 
signal macroparticles in each time bin. As the bunch population increases, the beam kick
increases the contribution to the SPU signal from photoelectrons produced with lower kinetic
energy.}
   \vspace{-1.1mm}
   \label{fig:icomp}
\end{figure}

The SPU signal during the first few nanoseconds following passage of a single bunch of 
positrons is produced primarily by photoelectrons 
produced on the bottom surface of the vacuum chamber, since it is the surface nearest the
detector where the electrostatic force from the beam bunch is directed at the detector. 
Figure~\ref{fig:icomp} 
clearly shows such signals increasing in magnitude
and arriving earlier as the bunch population is raised from \mbox{$1.6{\times}10^{10}$} to 
\mbox{$6.4{\times}10^{10}$}.  The increase in signal arises not only because of the higher rate
of synchrotron radiation photons, but also because of the greater beam kick accelerating
the lower-energy (and more common) photoelectrons into the detector sooner. Varying the
bunch current therefore allows a momentum analysis of photoelectron production. Note for example
that the leading edge of the pulse does not come appreciably earlier for higher bunch populations, 
indicating that the photoelectron energy distribution, rather than the beam kick, governs its
arrival time. Indeed, the 5-eV Gaussian function used in many prior ECLOUD simulations, for example
the {\cesrta} coherent tune shift models~\cite{ref:ipac10jac}, proved 
inadequate to model the SPU signals, since kinetic energies greater than 1~keV are required
to reproduce the leading edge of the SPU signals. The modeled signals shown in Fig.~\ref{fig:icomp}
were obtained by detailed tuning of the photoelectron energy distribution. It was found 
that a weighted superposition of two power-law functions reproduced the data as shown
for beam kicks ranging over a factor of four.

\begin{figure}[htbp]
   \centering
   \includegraphics*[width=78mm]{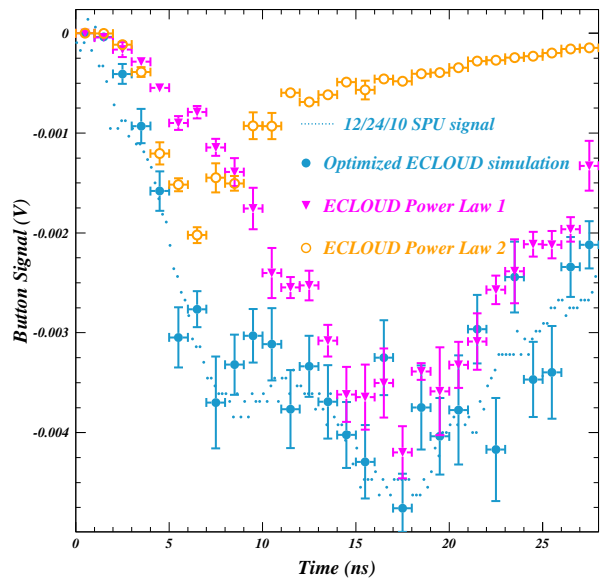}
   \caption{Contributions to the modeled signal from simulations including a single power
law contribution (orange circles show the higher energy contribution, pink triangles show 
the lower energy contribution) as well as the simulation including the weighted sum
of the two distributions which provides the observed degree of consistency with the measured
signal shape.}
   \vspace{-1.1mm}
   \label{fig:powerlaw}
\end{figure}

Figure~\ref{fig:powerlaw} shows an example of the individual 
contributions to the modeled signal of each of the two power laws, 
as well as the result of a simulation including both 
contributions.  The power laws are determined by the parameters $E_0$, $P_1$, and $P_2$ in the form 
$f(E_{\rm pe}) \propto E_{\rm pe}^{P_1}/(1+E_{\rm pe}/E_0)^{P_2}$.  The level of consistency with the measured signal
was obtained using a weight of 78\% for a photoelectron energy
distribution with peak energy 4~eV and $P_1=4$,
$P_2=6$ combined with a 22\% contribution from a distribution with peak energy 80~eV and
 $P_1=4$ and $P_2=8.4$.

Such detailed information on the photoelectron energy distribution applies only to photoelectrons
produced by photons which have scattered sufficiently often to reach the bottom of the vacuum
chamber. In contrast, the SPU measurements have not driven the need for such tuning
of photoelectron energies produced at the primary source point on the radially outward wall
of the vacuum chamber.
Consistency with the observed cloud buildup is obtained with low-energy photoelectron production,
such that the beam kick together with the intense space charge force due to the high concentration
of electrons at the primary synchrotron radiation impact point dominate the kinetic energy
distribution. Note also that the quantum efficiencies assumed for input parameters to the simulation
are averages over the incident photon energy distribution, which is different for photons
having undergone different numbers of reflections, so the ECLOUD input parameter definitions
were generalized to allow three independent quantum efficiency values for photoelectrons: 1)~those produced
at the primary synchrotron radiation impact point
on the radially outward side of the beam-pipe, 2)~those produced at the point on the 
inward side of the beam-pipe opposite the primary impact point,
where most absorbed photons have undergone a single reflection, and 3)~those produced elsewhere. 
Finally, we 
checked that the introduction of high-energy components in the photoelectron energy
distributions did not affect the level of agreement obtained previously with the 
coherent tune measurements, i.e. the tune shift modeling is insensitive to 
the photoelectron energy distribution for scattered photons.

\section{Secondary Production Model}

The SPU signal from a witness bunch provides sensitivity to the development of the cloud
produced by the first bunch, including secondary electron production. The witness signal
includes contributions from cloud electrons in the region between the beam and
the bottom of the vacuum chamber, and so will be greater in magnitude and earlier than
the signal from the leading bunch. The cloud population in the region of signal sensitivity
depends directly on the kinetic energy of the cloud electrons, both the photoelectrons and
the secondary electrons. At a time in the cloud development when a substantial fraction of the
cloud electrons have undergone a single wall interaction, the energy distribution will
be sensitive to the production energies of the true secondaries. At later times the cloud energy
distribution is stabilized by the predominance of elastic wall interactions.

\begin{figure}[htbp]
   \centering
   \includegraphics*[width=40mm]{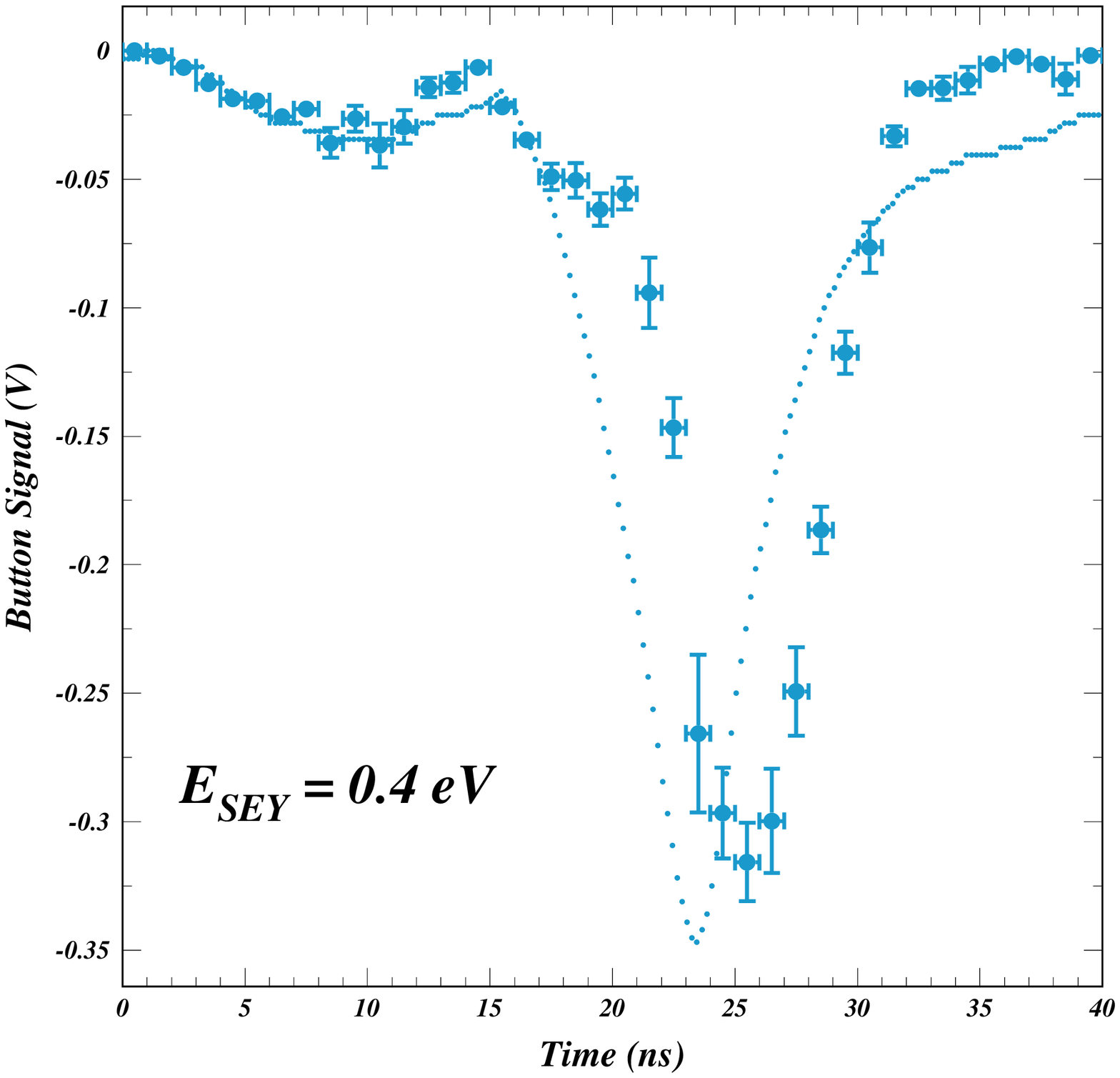}
   \includegraphics*[width=40mm]{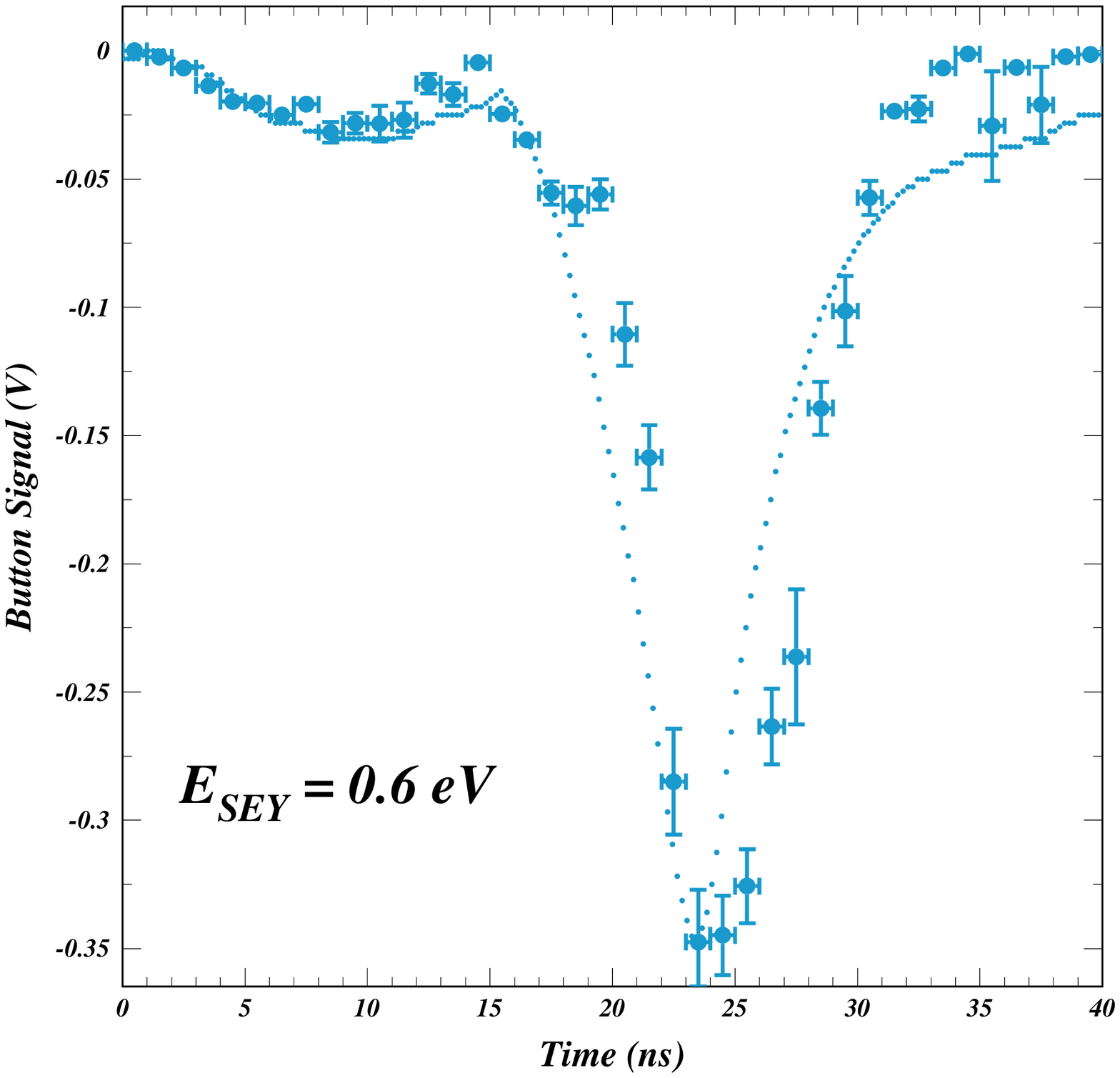}\\
   \includegraphics*[width=40mm]{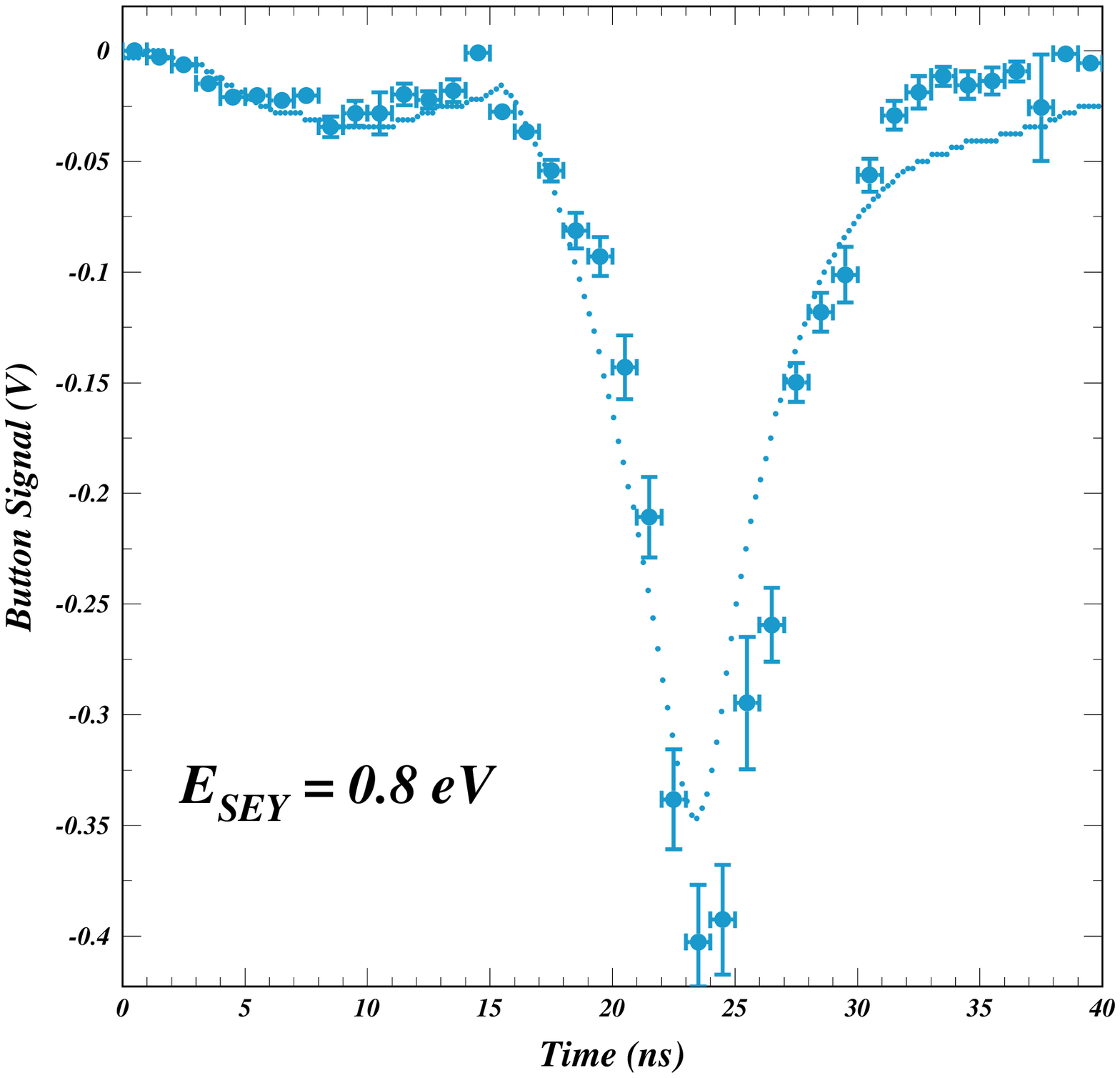}
   \includegraphics*[width=40mm]{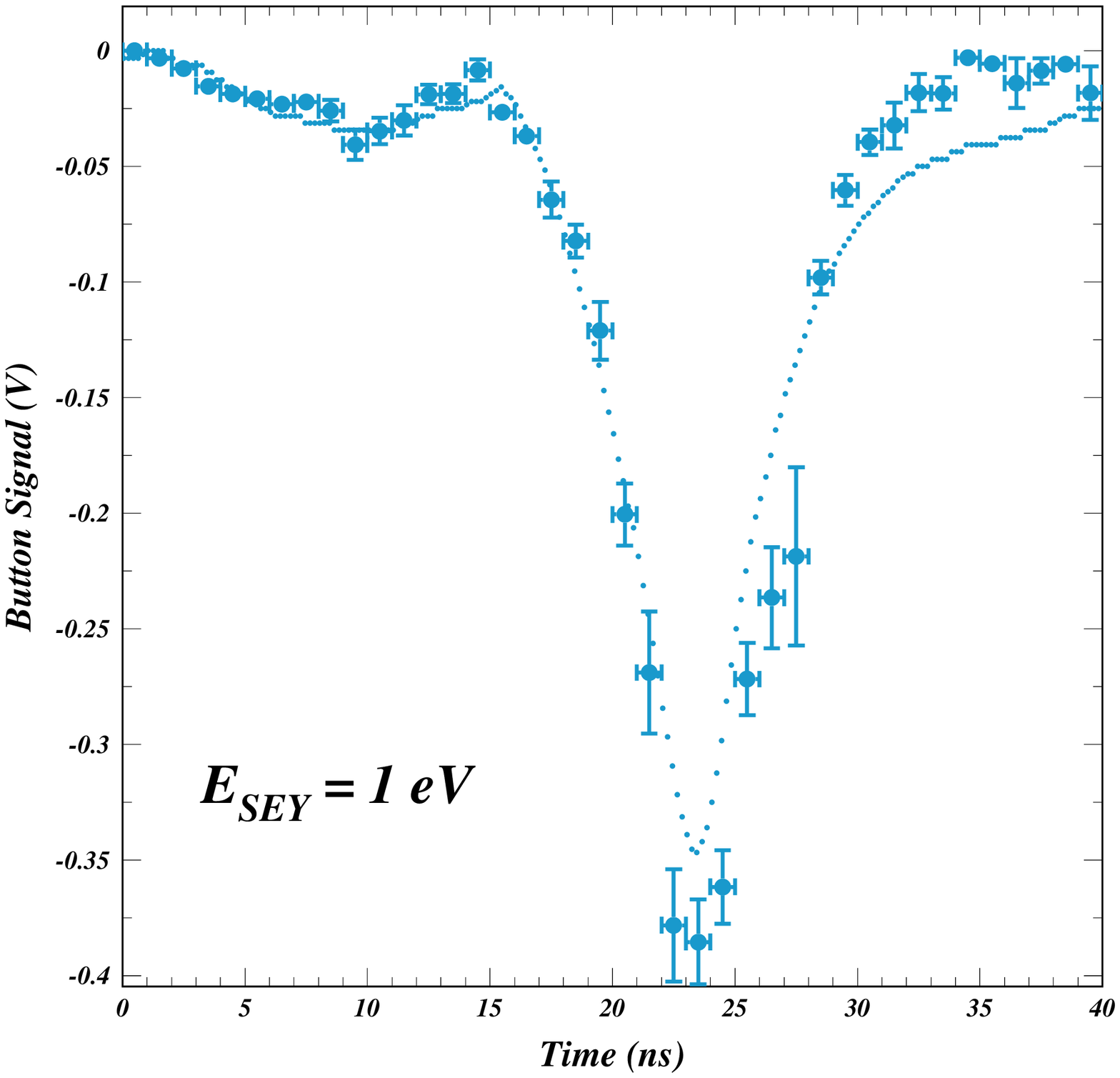}
   \caption{SPU signal for two 5.3~GeV 
positron bunches of population \mbox{$8{\times}10^{10}$} spaced by 28~ns.
Each of the four figures show the result of a simulation with a different assumption for
the secondary emission model parameter $E_{\rm SEY}$, which determines the production kinetic
energy distribution for electrons produced via the true secondary emission process in 
interactions with the vacuum chamber surface. The sensitivity to the cloud position and energy
distribution at the time of the passage of the witness bunch provides a lower limit
on this parameter with a sensitivity better than 0.2~eV.}
   \vspace{-1.1mm}
   \label{fig:semax}
\end{figure}

Figure~\ref{fig:semax} shows an example of the model sensitivity to the
true secondary energy distribution. Two 5.3~GeV
positron bunches of population \mbox{$8{\times}10^{10}$}
separated by 28 ns provided the signal shown in the four figures, each of which also shows
the ECLOUD model with differing assumptions for the true secondary production energy distribution.
The distribution is parameterized as $f(E_{\rm sec}) \propto E_{\rm sec} \; {\rm exp}(-E_{\rm sec}/E_{\rm SEY})$. 
The arrival time of the signal from the witness bunch sets a lower bound on the assumed value
of $E_{\rm SEY}$ with a sensitivity better than 0.2~eV. 

We also found an upper limit on the parameter $E_{\rm SEY}$ to be imposed by the late tail 
(40-80~ns after the bunch passage) of a single bunch SPU signal. Values below  
$E_{\rm SEY} = 1.2$~eV resulted in a broad late tail
from photoelectrons produced on the outside of the vacuum chamber which is not observed
in the measured signals. Since the sensitivity was again found to be about 0.2~eV, 
these two phenomena provide a remarkably tight constraint on the production energy
distribution for true secondary electrons.

\begin{figure}[htbp]
   \centering
   \includegraphics*[width=78mm]{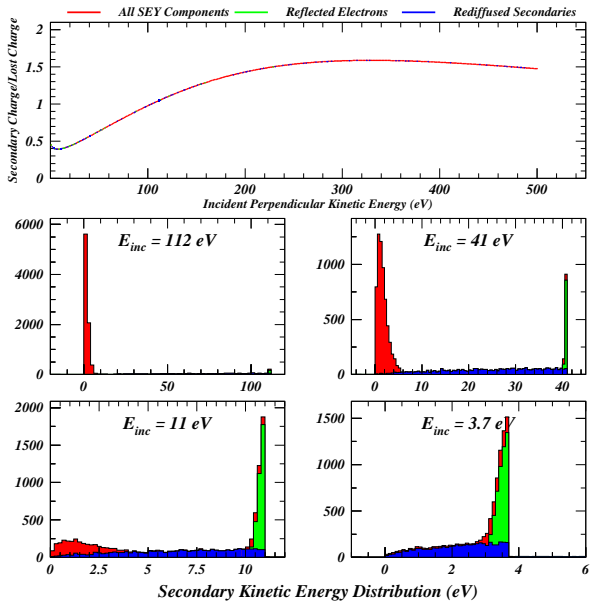}
   \caption{
Example of the secondary emission model implemented in the ECLOUD code. The top plot shows
the total yield curve including the contributions from the 
true secondary, re-diffused and elastic processes. The lower four plots show the 
energy distributions of the secondaries in the case of perpendicular incidence for 
electrons of energy 112, 41, 11 and 3.7~eV.}
   \vspace{-1.1mm}
   \label{fig:seychar}
\end{figure}

Such measurements of the time dependence of EC development afford discriminating
power between the three components of the secondary yield model, since they are sensitive to
the energy distribution in the cloud, and therefore to the relative probabilities of the types
of wall interaction, each of which produces secondaries of characteristic energies. Figure~\ref{fig:seychar}
shows a typical example of the yield curve and emitted energy distributions of
the secondary emission model implemented in ECLOUD. 
The secondary yield values in this example are typical of uncoated aluminum: 1.4, 0.2, and 0.4
for the true secondary, re-diffused and elastic processes, respectively. At low incident energy,
the elastic process dominates, while the re-diffused contribution is independent of energy for
energies greater than a few~eV. The true secondary process dominates at high incident energy,
and produces secondaries carrying only a few~eV. The model includes an RMS smearing of 0.3~eV
for the outgoing energies of the elastics. For purposes of comparison to the discussion in 
Ref.~\cite{ref:prl93_nr1} the secondary electron energies are shown for incident energies of
112, 41, 11 and 3.7~eV.

We have found
that the relative rate of re-diffused secondaries is constrained by the lifetime of the cloud
produced by a single
bunch as manifested in the tail of the SPU signal~\cite{ref:ibic12sikora}. A model excluding
the re-diffused process underestimates the EC lifetime from a single bunch with a sensitivity
better than 10\%. Measurements on a bare aluminum vacuum chamber were best matched with
a re-diffused contribution of 20\%, in quantitative agreement with the constraint provided
by models of the {\cesrta} coherent tune shift data, where the omission of the re-diffused
component resulted in an underestimate of the vertical tune shift increase
in a strong dipole magnetic field along a train of 45 2.1~GeV bunches each 
carrying \mbox{$1.3{\times}10^{10}$} positrons~\cite{ref:ipac10jac}.

\section{Beam Conditioning Effects}
The good reproducibility of the SPU measurements on a time scale of months has provided the
ability to determine details of the beam conditioning process by observing the 
long-term time dependence
of optimized model parameters. These studies are a subset of the {\em in situ} vacuum chamber
comparisons in the same radiation environment (i.e. the same place in the CESR ring) for
the case that the same chamber was left in place. 
The two 
regions in CESR equipped with SPU detectors 
differ in radiation environment, since the dominant source points are in dipole magnets of differing
strengths. At 5.3~{\gev}, for example, the source dipole field is 3~kG (2~kG) in the west (east) region
for a positron beam, resulting in a critical energy of 5.6~{\kev} (3.8~{\kev}). In addition, the 
distribution of reflected photons differs. By comparing SPU signals recorded at the same place
in the ring with the same beam energy, bunch spacing and bunch population, many systematic contributions to
the comparisons are avoided, and relatively simple changes to the modeling suffice to quantify the different
properties of the vacuum chambers.

\begin{figure}[b]
   \centering
   \includegraphics*[width=78mm]{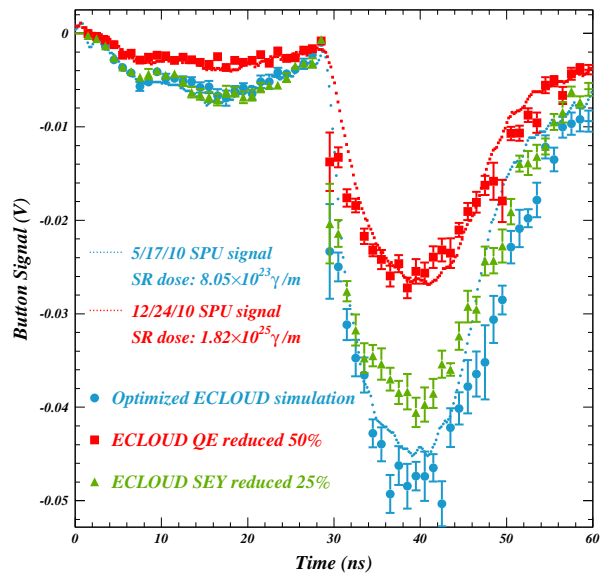}
    \caption{SPU signals measured in an a-C-coated chamber in May (blue dotted line) 
and December (red dotted line) of 2010 for two 5.3~{\gev}, 28-ns-spaced bunches each
carrying \mbox{$4.8{\times}10^{10}$} 
positrons. The ECLOUD model 
optimized for the May data is shown as blue circles, the error bars showing the 
signal macroparticle statistical uncertainties. 
The conditioning effect due to an exposure to synchrotron radiation 
increased by a factor of about twenty is reproduced by a 50\% decrease in the modeled 
quantum efficiency for photoelectron production (red boxes). A reduction in the secondary 
yield of 25\% (green triangles) is inconsistent with the observed effect.
}
   \vspace{-1.1mm}
   \label{fig:acarbon1}
\end{figure}

The first example of beam conditioning effects we studied was the
case of an amorphous-carbon-coated aluminum chamber.  Figure~\ref{fig:acarbon1}
shows signals recorded
in May and December 2010 for two 5.3~{\gev} 28-ns-spaced bunches each carrying \mbox{$4.8{\times}10^{10}$} 
positrons, corresponding to a bunch current of 3~mA. 
During the intervening time interval, 
CESR had operated at high current as an X-ray research facility, with the 
consequence that synchrotron 
radiation dose on the chamber had increased by a factor of about 20, from 
\mbox{$8.05{\times}10^{23}$} to  \mbox{$1.82{\times}10^{25}$~$\gamma$/m}. 
Also shown is the ECLOUD model optimized to reproduce the May measurement. 
Since conditioning affects the signals following each bunch similarly, we can conclude that the 
change is in the quantum efficiency rather than in the secondary yield. The December measurement is 
reproduced by a 50\% decrease in the modeled quantum efficiency for photoelectron production. 
A reduction in the secondary yield of 25\% is inconsistent with the observed effect, since
the modeled leading bunch signal remains unchanged while the measured signal is clearly
reduced.

In order to investigate the conditioning process for a chamber which
had not seen any beam at all, we installed such a chamber in September, 2011, recording
SPU measurements as soon as beam operations began. These measurements were then compared to
measurements made in November. The synchrotron radiation dose  between the two measurements increased from 
\mbox{$4.53{\times}10^{20}$} to  \mbox{$6.23{\times}10^{24}$~$\gamma$/m}, corresponding
to an integrated beam dose increase of about \mbox{$2{\times}10^{-2}$} to \mbox{$4{\times}10^{2}$~Amp-hours}.

\begin{figure}[b]
    \centering
    \includegraphics*[width=78mm]{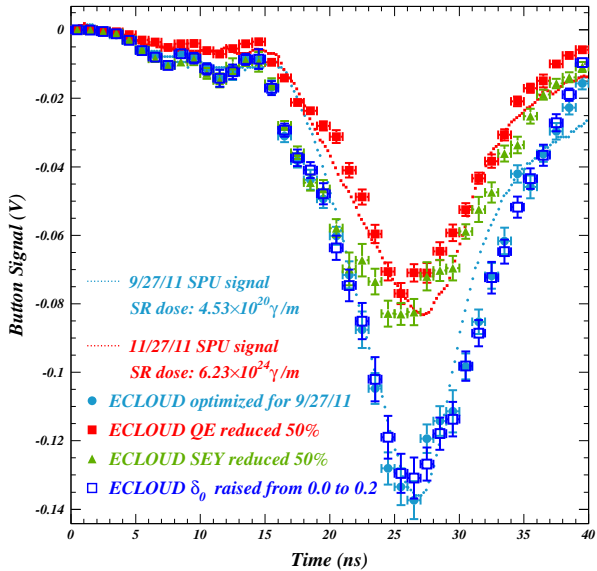}
    \caption{SPU signals measured in an a-C-coated chamber in September (blue dotted line) 
and November (red dotted line) of 2011 for two 5.3~{\gev}, 14-ns-spaced bunches each
carrying \mbox{$4.8{\times}10^{10}$} 
positrons. The ECLOUD model 
optimized for the September data is shown as solid cyan circles. The red squares show the results of a model
in which the quantum efficiency has been reduced 50\%, matching the November data reasonably well. 
The green triangles show the result of a simulation
in which the peak secondary yield value is reduced 50\%. The open blue squares show the effect of raising
the elastic yield value $\delta_0$ from 0\% to 20\%.
}
   \vspace{-1.1mm}
    \label{fig:acarbon2}
\end{figure}

Figure~\ref{fig:acarbon2} shows signals recorded with two 5.3~{\gev} 14-ns-spaced bunches each carrying \mbox{$4.8{\times}10^{10}$} 
positrons, corresponding to a bunch current of 3~mA.  
Between the
two measurements the photon dose increased from \mbox{$4.53{\times}10^{20}$} to  \mbox{$6.23{\times}10^{24}$~$\gamma$/m}.
Also shown is the ECLOUD model optimized to reproduce the September measurement.
The November measurement is reproduced by a 50\% decrease in the modeled quantum for photoelectron production. 
A reduction in the SEY of 50\% is inconsistent with the observed effect, since
the modeled leading bunch signal remains unchanged. Thus we conclude that the 
early conditioning process is similar to one previously measured in a well-conditioned chamber.

\begin{figure}[htbp]
    \centering
   \includegraphics*[width=78mm]{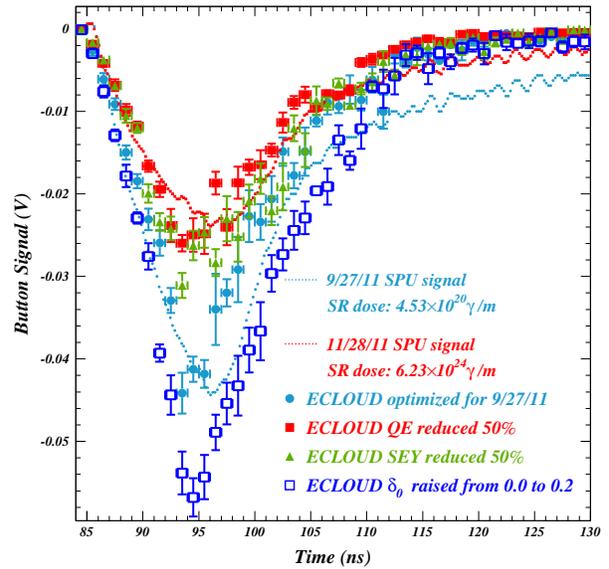}
    \caption{
SPU signals and modeling for a two-bunch signal with 84-ns spacing
showing the sensitivity to the elastic secondary yield component
in the time-resolved measurement technique, largely uncorrelated to
the sensitivity to the other two secondary emission processes.
}
   \vspace{-1.1mm}
    \label{fig:elasticyield}
\end{figure}

Figure~\ref{fig:acarbon2} also shows the results of a model in which the yield value 
$\delta_0$ for the elastic component of the secondary
yield has been increased from 0\% to 20\%. The modeled 14-ns signal is insensitive to such a 
change in the elastic yield.
In contrast, the two-bunch signals for the case of \mbox{84-ns} separation shown in
Fig.~\ref{fig:elasticyield} 
clearly show sensitivity
to the elastic yield component, and exclude a value as high as 20\%. Such a comparison 
permits the conclusion
that the measurements are inconsistent with a conditioning effect in the elastic yield of 20\%.
Such low values for the elastic yield are characteristic of the amorphous carbon, diamond-like carbon,
and TiN coatings, contrasting with a value
closer to 50\% required to match the SPU data for an uncoated aluminum chamber~\cite{ref:ecloud10jac},
as discussed in the next section.

\section{EC Lifetime Studies}

While the awareness of the sensitivity of the SPU measurements to
the parameters of photoelectron production was largely motivated by inadequacies
of the model discovered in its application to recent measurements, the original
intended use of these time-resolved cloud measurements was to provide a quantitative 
estimate of the elastic yield parameter in the secondary electron yield model. 
A similar investigation was performed at RHIC~\cite{ref:epac2006gr}.
The basic concept is that the mature cloud long after passage of any beam bunch
is dominated by low-energy electrons
which undergo primarily elastic interactions with the vacuum chamber wall.

\begin{figure}[htbp]
   \centering
   \includegraphics*[width=78mm]{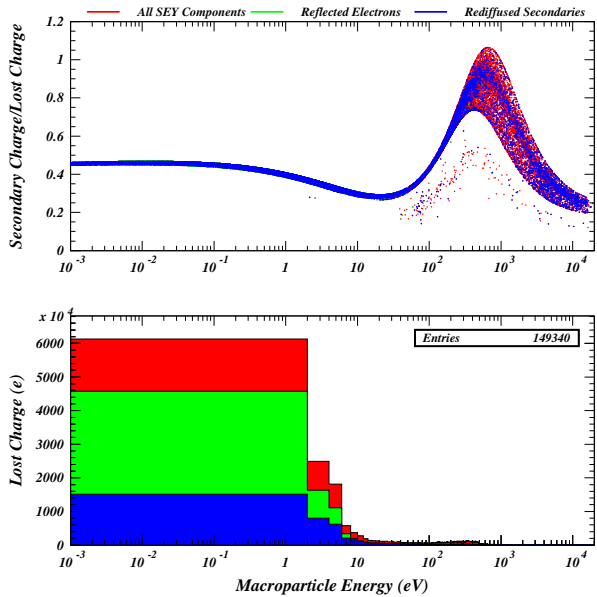}
   \caption{Secondary yield population curve typical of the ECLOUD model for the SPU 
signals.The upper plot shows the yield value (ratio of secondary macroparticle
charge to that of the incident charge) as a function of the incident kinetic energy. 
The lower plot shows the incident
energy distribution. The elastic and re-diffused components are shown in green and blue, respectively.
The sum of all three components, true, elastic and re-diffused, is shown in red. 
The three colors are plotted on top of each other, so the upper plot shows primarily 
blue at low energy, even though the elastic process dominates, as shown in the lower plot.
}
   \vspace{-1.1mm}
   \label{fig:seypop}
\end{figure}

Figure~\ref{fig:seypop} shows an ECLOUD secondary yield population curve typical
of the  signal simulations for a carbon- or TiN-coated aluminum
vacuum chamber.
The true secondary yield maximum at
400~{\ev} ranges from a minimum of 0.8 to a maximum of 1.1 owing to the dependence on
incident angle. At low energy the yield value is dominated by the elastic interactions 
with the chamber wall.
This case exhibits a total yield at low energy of 45\%, of which 40\% is elastic. Under the true
secondary peak some cases are shown where ECLOUD generates two secondaries each carrying half
the secondary charge in order to limit the maximum charge of a macroparticle.

\begin{figure*}[htbp]
   \centering
   \includegraphics*[width=41mm]{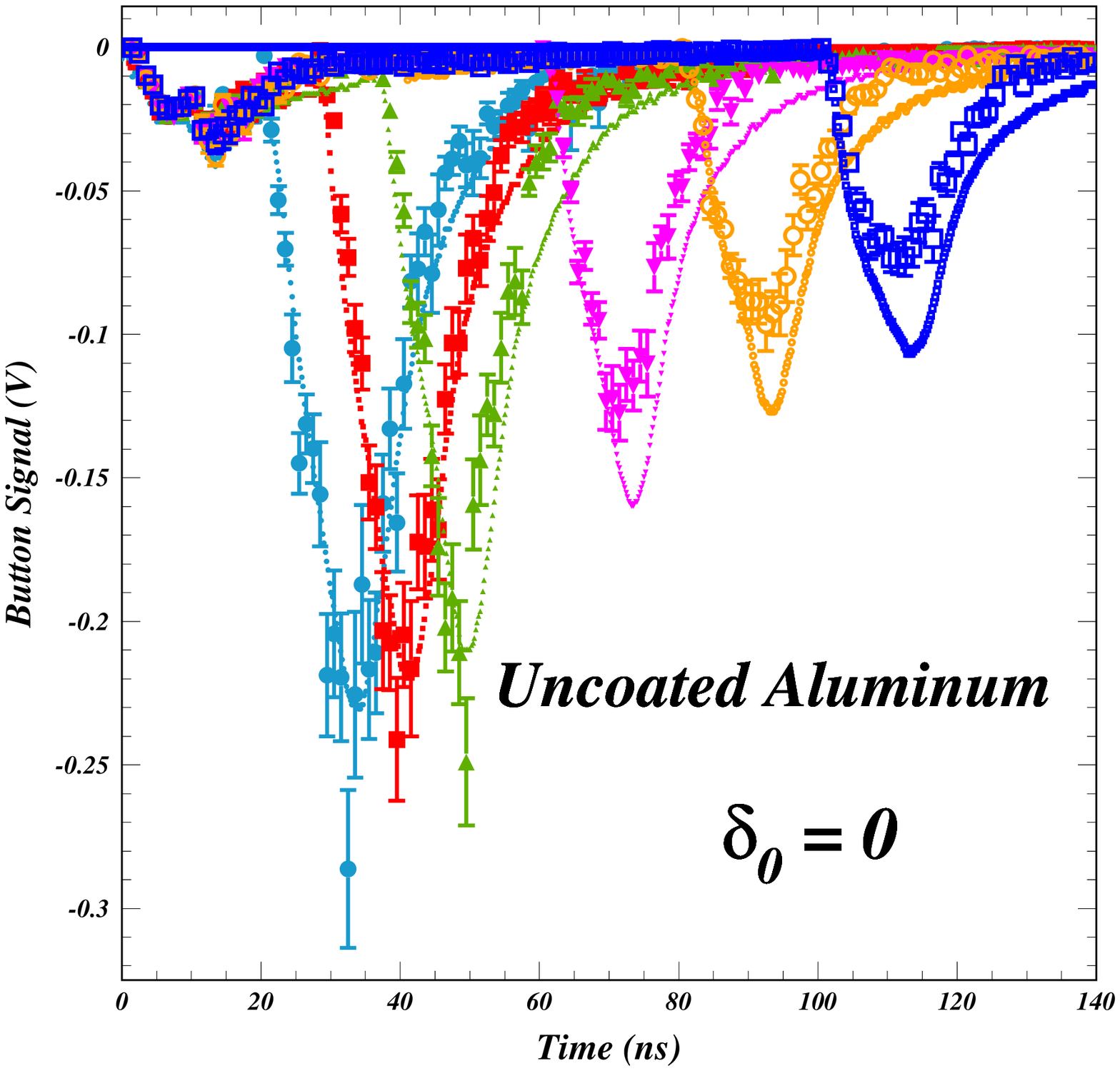}
   \includegraphics*[width=41mm]{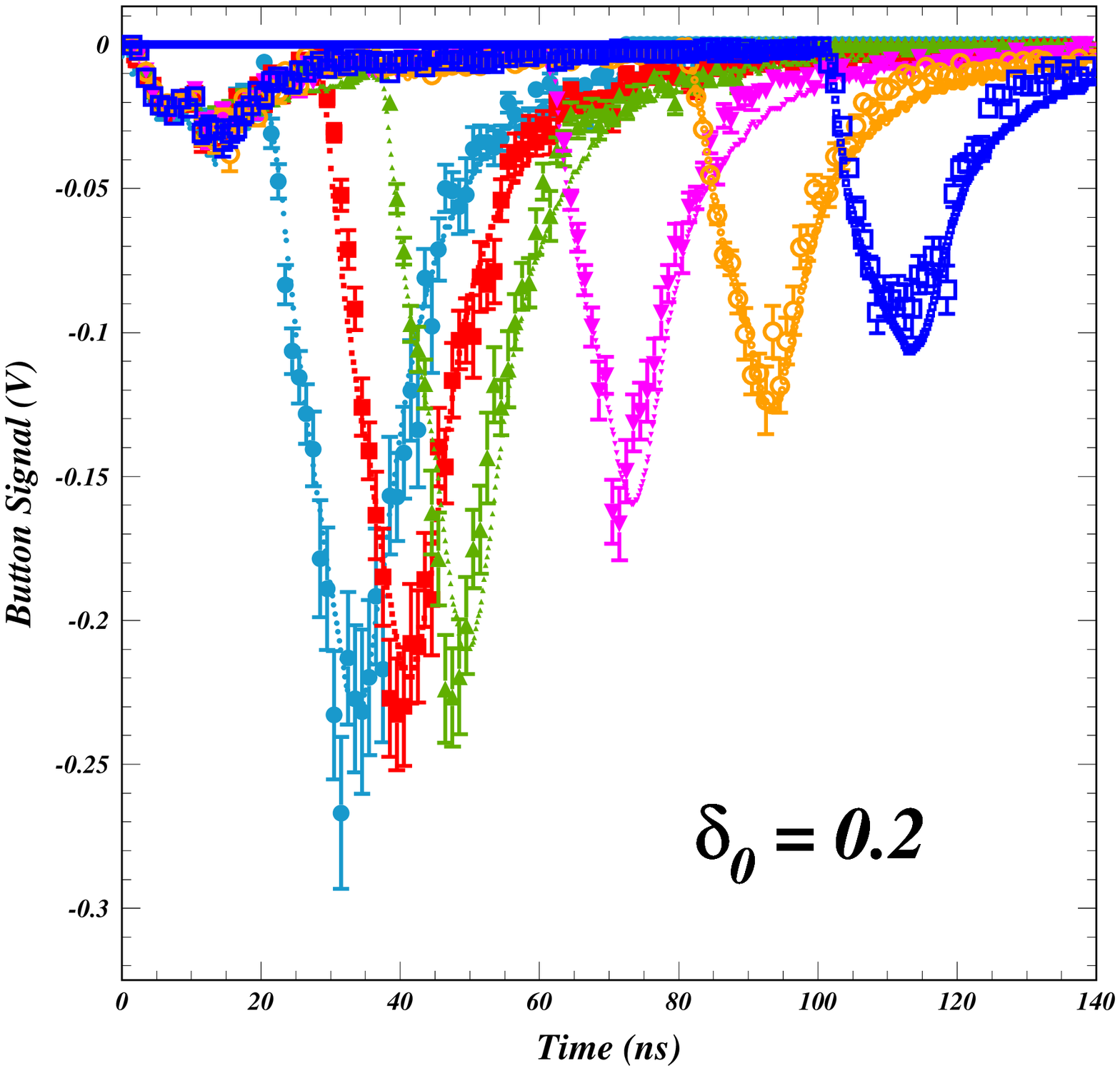}
   \includegraphics*[width=41mm]{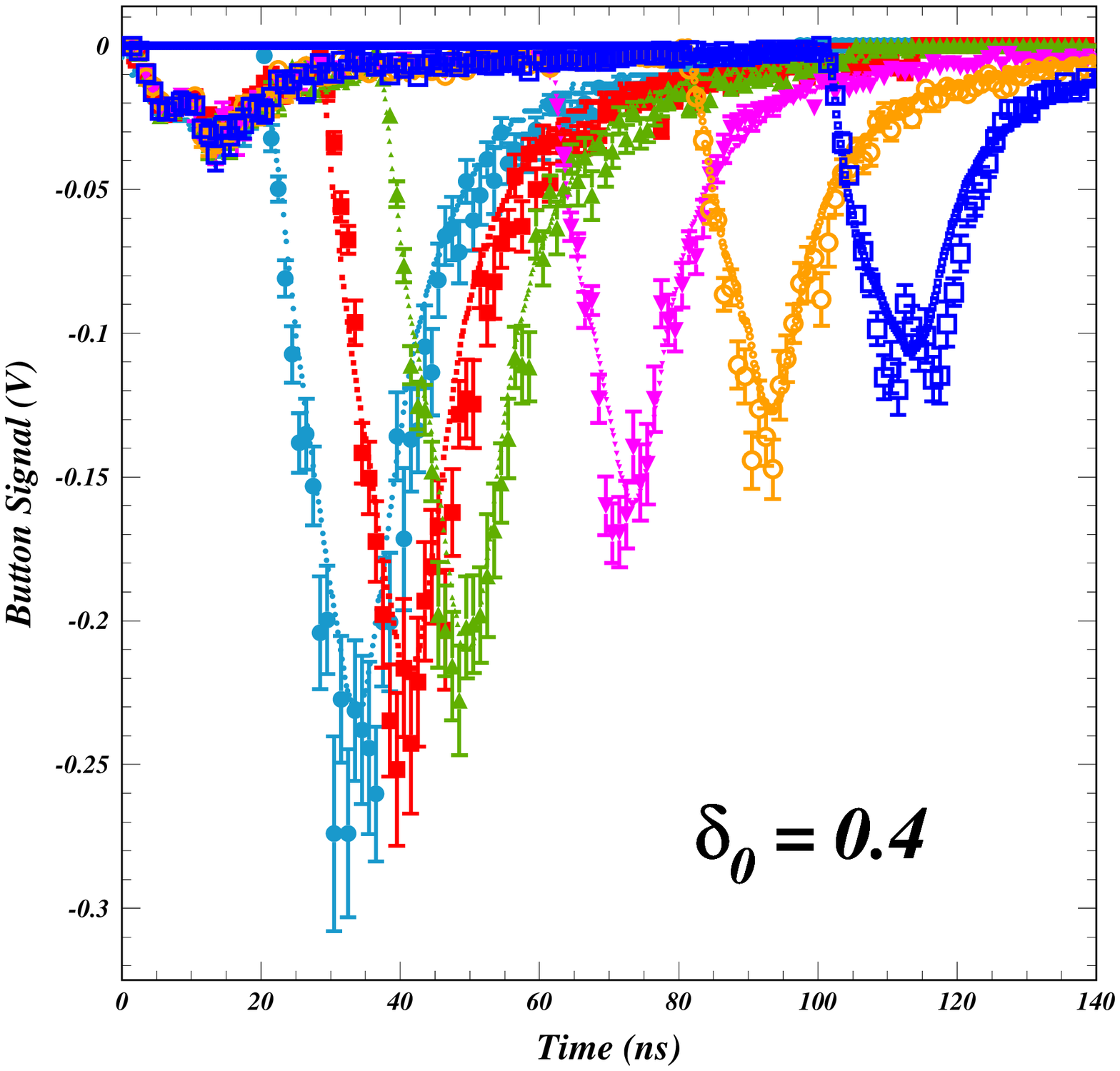}
   \includegraphics*[width=41mm]{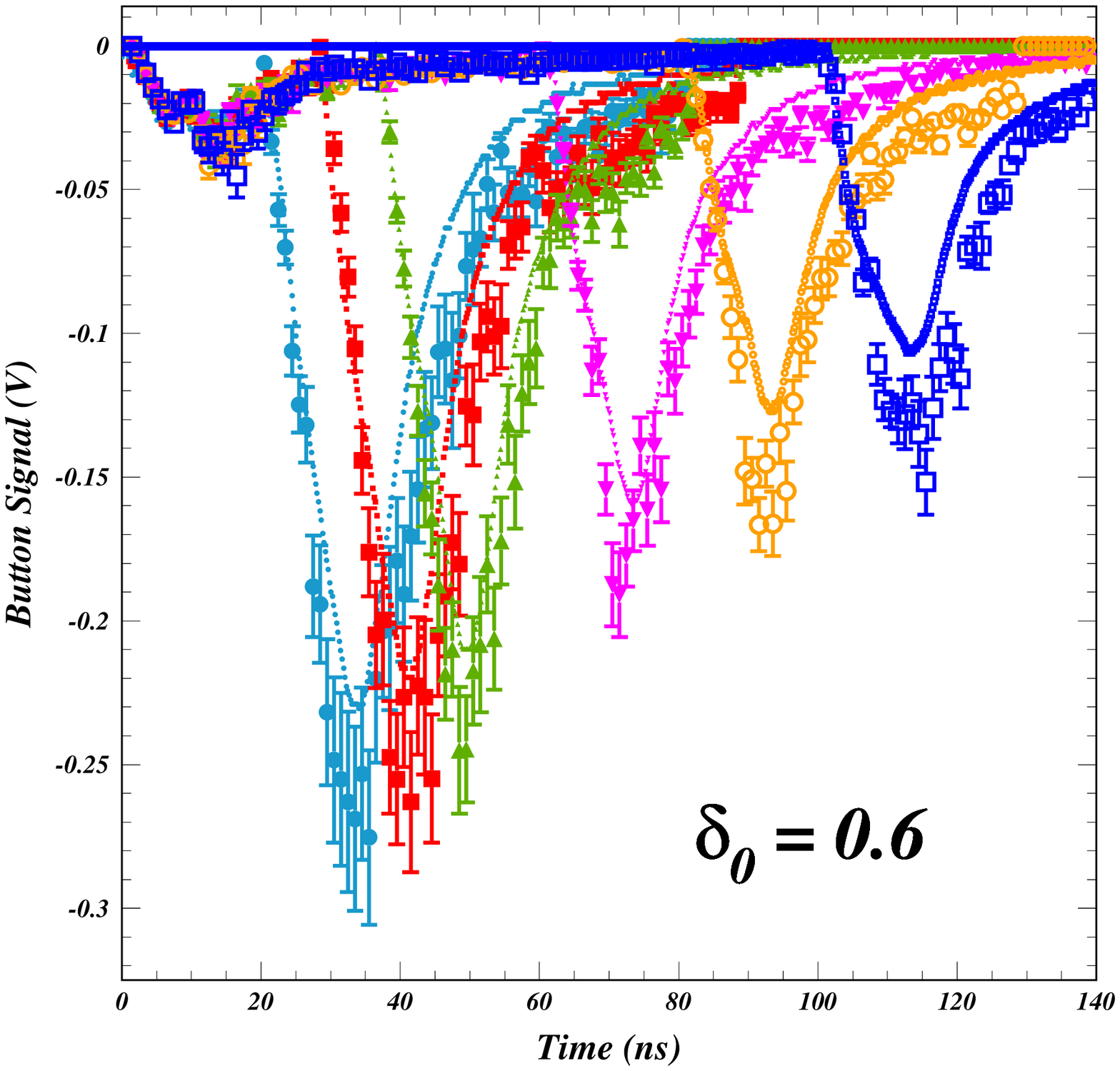}\\
   \includegraphics*[width=41mm]{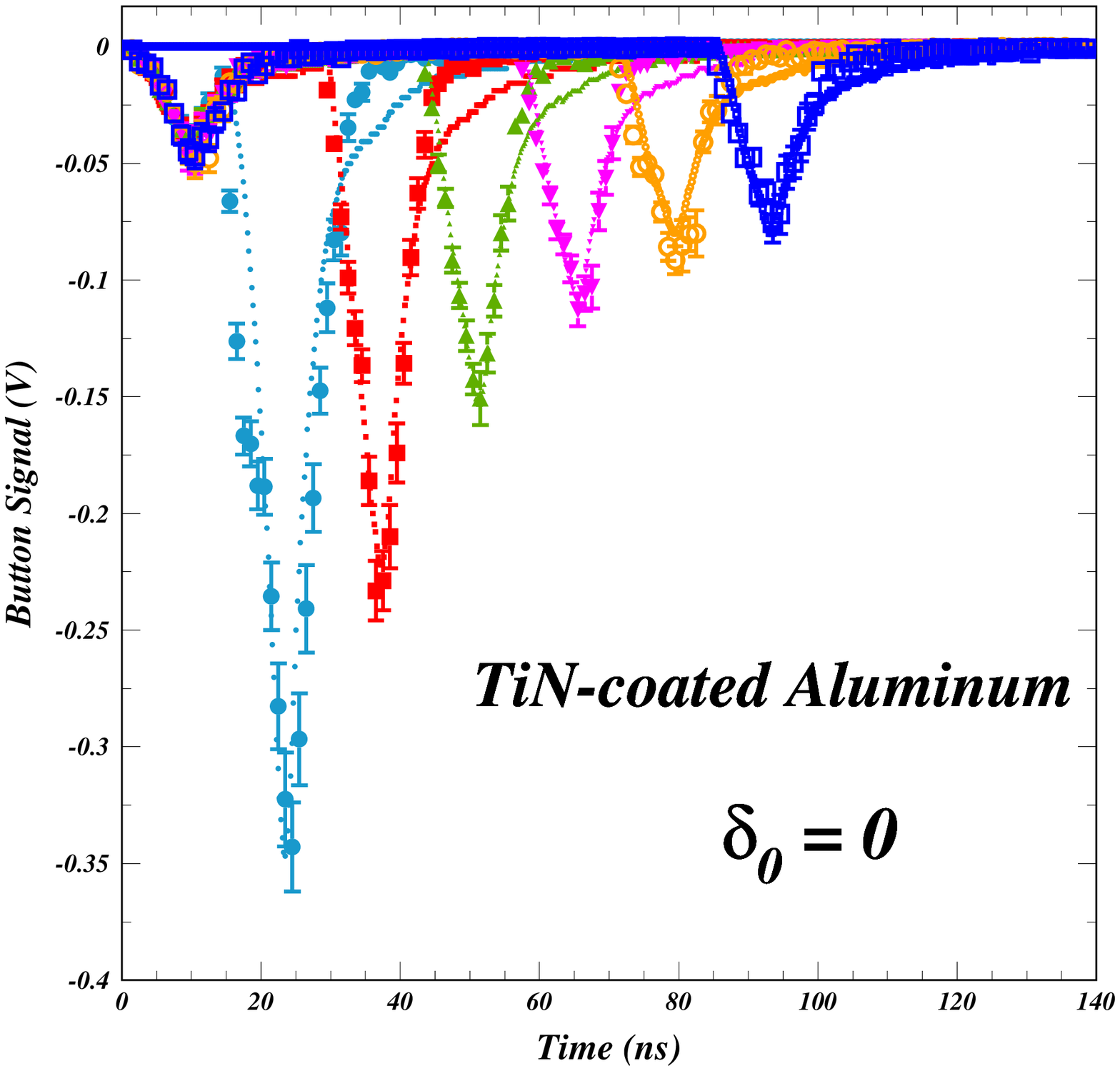}
   \includegraphics*[width=41mm]{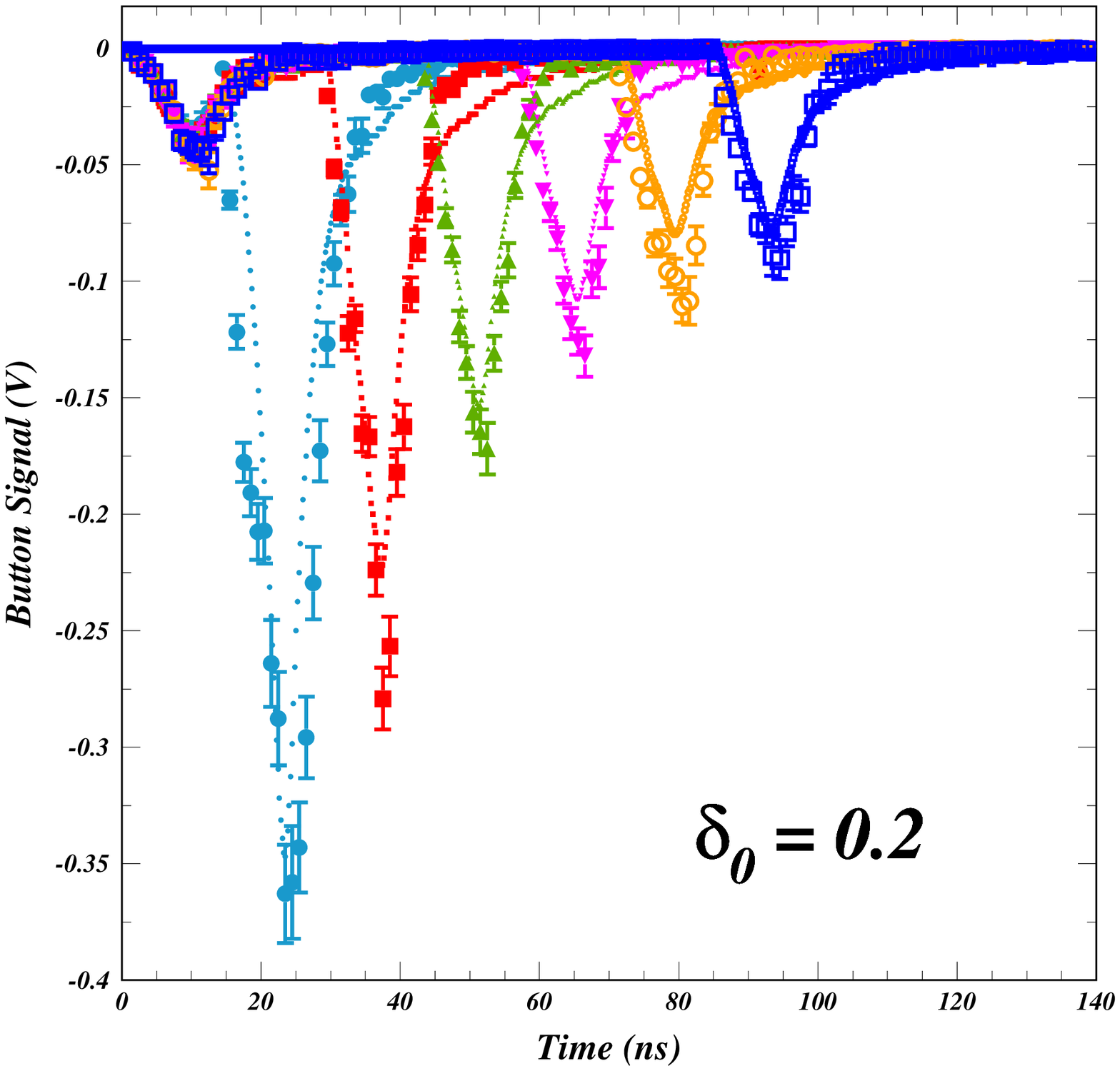}
   \includegraphics*[width=41mm]{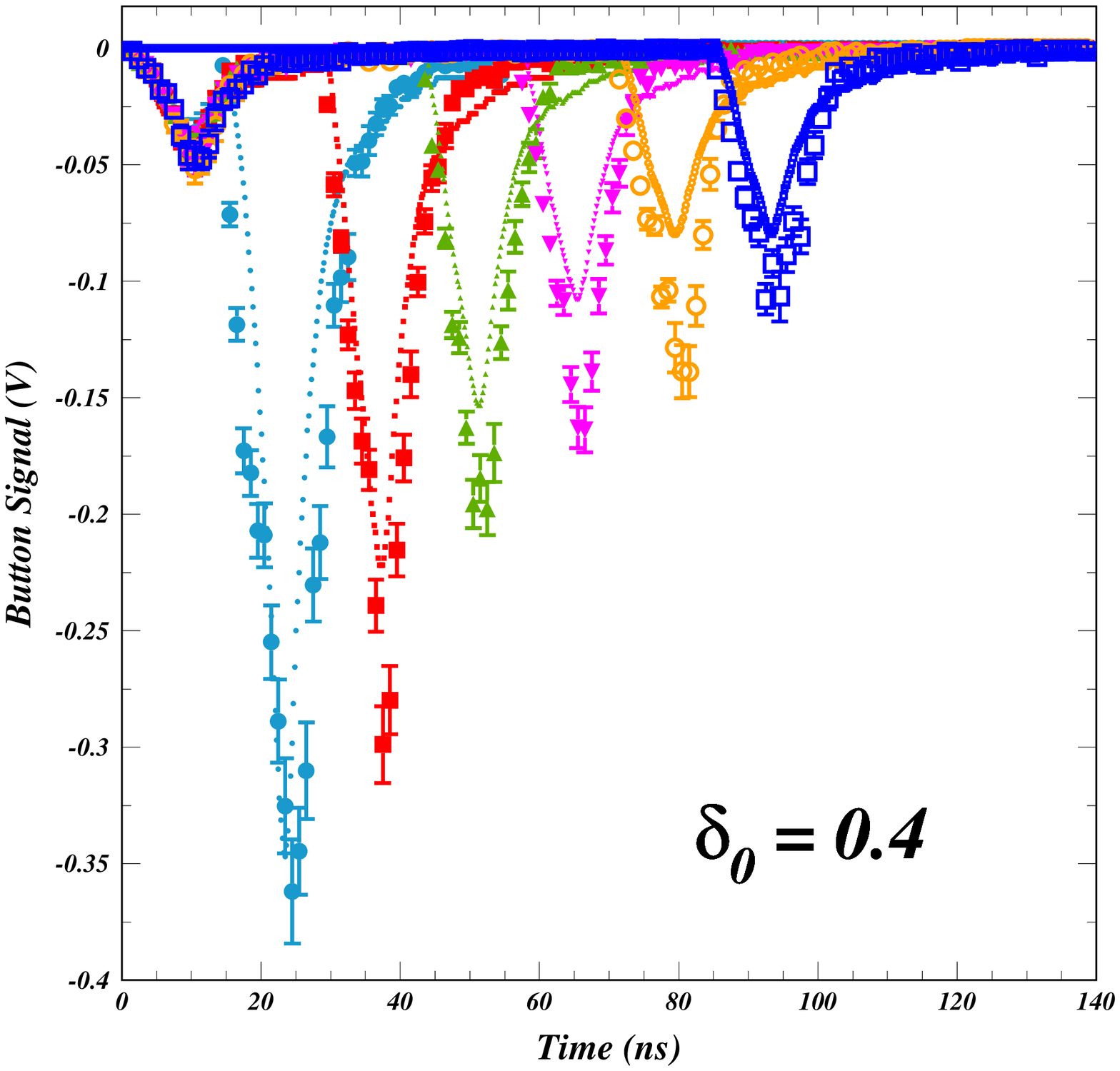}
   \includegraphics*[width=41mm]{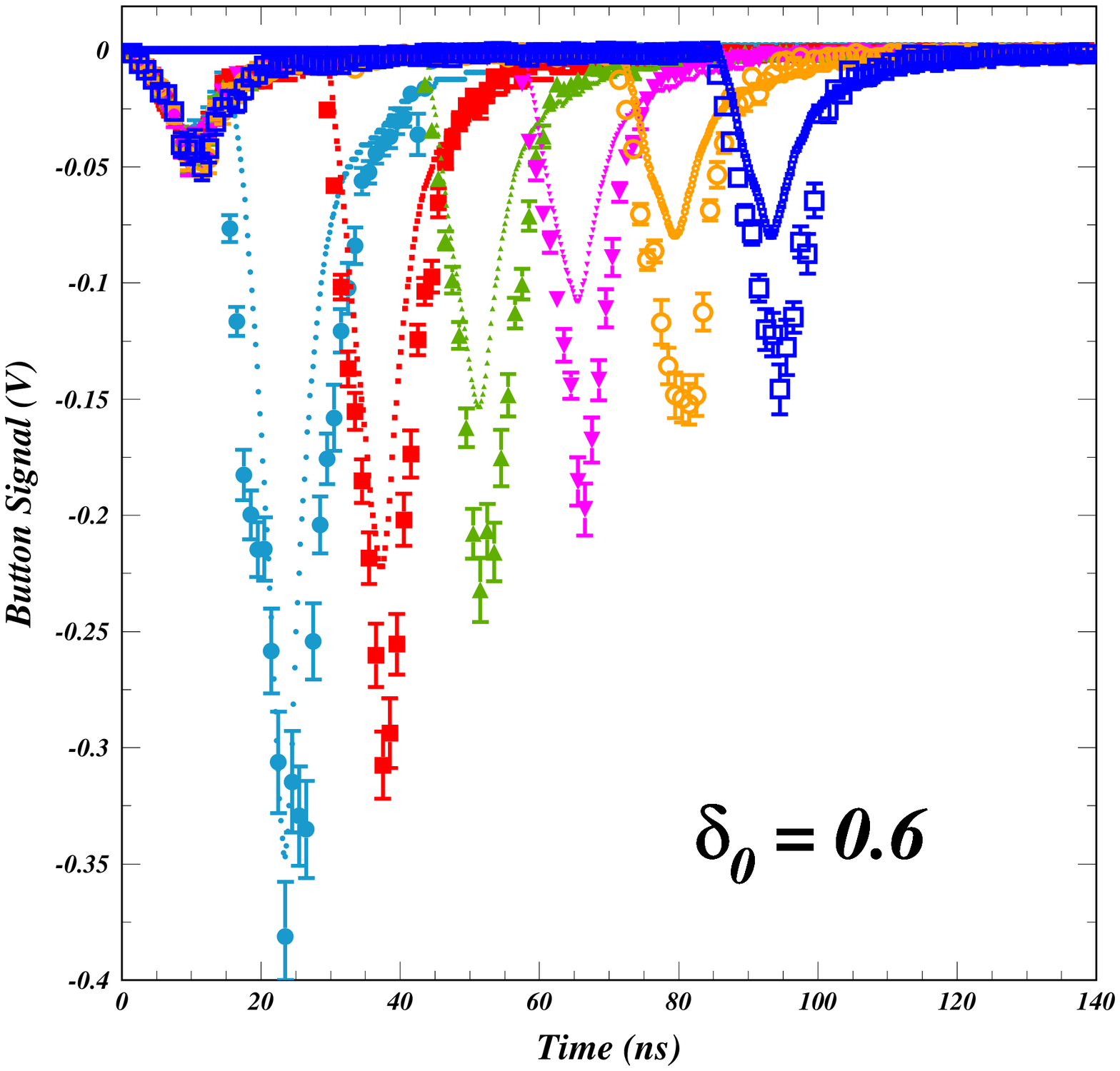}
   \caption{
Witness bunch study constraining the model parameter for the elastic yield ${\delta}_0$. The upper row 
compares model results to six superposed two-bunch SPU signals produced by 5.3~GeV positron bunches of population 
\mbox{$4.8{\times}10^{10}$} in an uncoated aluminum vacuum chamber. The bottom row shows the equivalent comparison
for positron bunches of population \mbox{$8.0{\times}10^{10}$} in a TiN-coated aluminum chamber. The value of the model input parameter
for the elastic yield is raised from 0 to 60\% from the first to the last column, exhibiting an optimal value about about 40\%
for uncoated aluminum and less than 20\% for the TiN-coated chamber surface.}
   \vspace{-1.1mm}
   \label{fig:delta0}
\end{figure*}

Figure~\ref{fig:delta0} shows how the witness bunch studies constrain the model parameter
for the elastic yield $\delta_0$. 
The upper row shows a scan of the modeled $\delta_0$
parameter for six two-bunch SPU signals with spacings of 20, 24, 36, 60, 80 and 100 ns.
The two 5.3 GeV positron bunches each carry a population of \mbox{$4.8{\times}10^{10}$}. These
signals were recorded in an uncoated aluminum chamber. The lower row shows a similar
study for a TiN-coated chamber. In this case the each of the two positron bunches carries
a population of \mbox{$8.0{\times}10^{10}$} and the spacings are 14, 28, 42, 56, 70 and 84 ns. 
The witness signals with longer delays between bunches clearly provide good discriminating 
power for the elastic yield, showing a sensitivity somewhat better than 20\%. The optimal
value of $\delta_0$ for the uncoated aluminum chamber is about 40\%, consistent with the
value of 50\% used in the simulations which successfully modeled the {\cesrta} coherent
tune shift measurements. In contrast, these witness-bunch measurements for a TiN-coated chamber 
exclude values for $\delta_0$ greater than 20\%.

\section{Installation of Time-resolved Retarding Field Analyzers}

\begin{figure*}[htbp]
   \centering
   \includegraphics*[width=160mm]{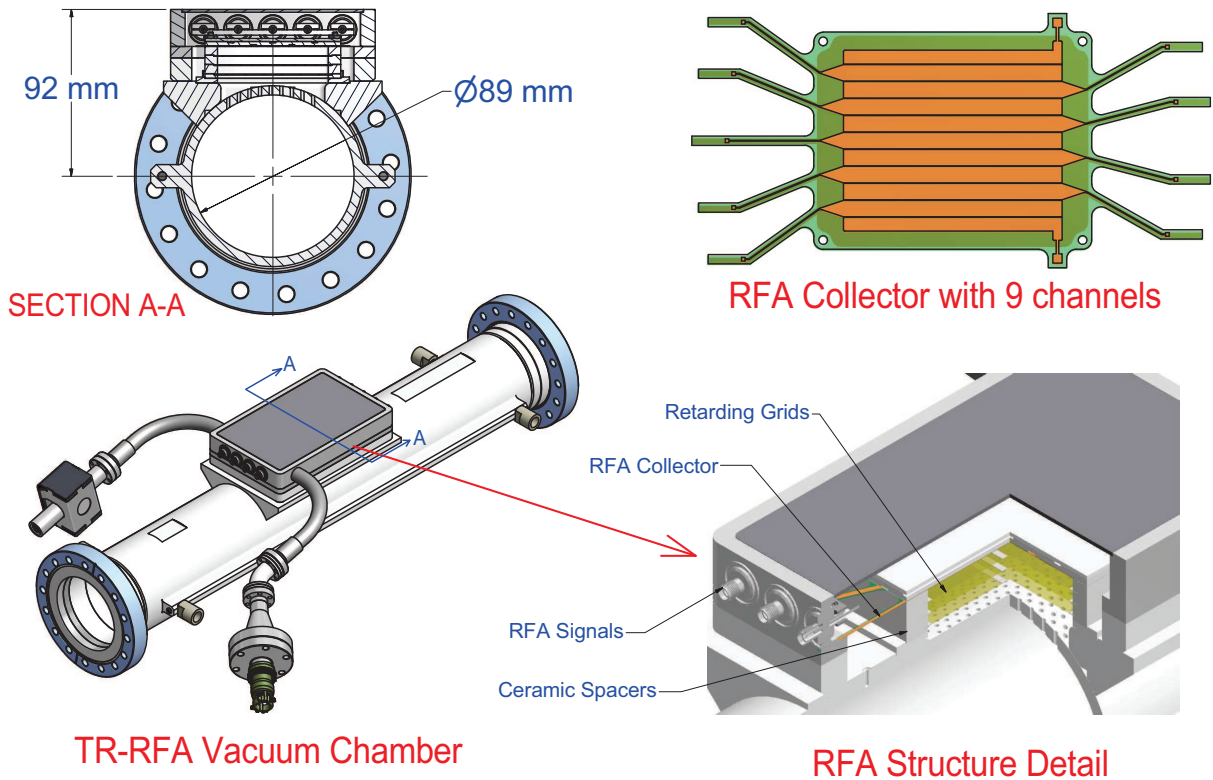}
   \caption{Engineering drawings of a {\cesrta} time-resolved retarding field analyzer. The two drawings in the left 
column show the 8.9-cm-diameter custom
vacuum chamber on which the TR-RFAs are installed. The nine collectors are arranged as shown on the upper right. 
The interior grid structure is shown on the lower right.}
   \vspace{-1.1mm}
   \label{fig:tr_rfa}
\end{figure*}

Time-resolved RFAs (TR-RFA) combine the properties of SPUs and retarding-field analyzers by providing signals 
with time resolution of about a nanosecond
from a finely segmented collector array and
a grid which can be biased to define variable sensitivity to cloud electron energy.
Figure~\ref{fig:tr_rfa} shows the design of four TR-RFAs installed in the CESR ring in four custom vacuum chambers, each of which is located at the center
of a chicane dipole magnet which can apply a field as high as 800~G.
The middle grid is used to provide the retarding field. There are nine collectors etched on a kapton flex circuit, each connected to an
SMA feed-through. The collectors are biased at +50~V to prevent secondary electrons from leaving the collector surface.
The four chamber designs are chosen to study the EC mitigation techniques proposed for the ILC positron damping ring: TiN-coated and bare aluminum, each type
with and without grooved lower and upper surfaces.  

\begin{figure}[htbp]
   \centering
   \includegraphics*[width=78mm]{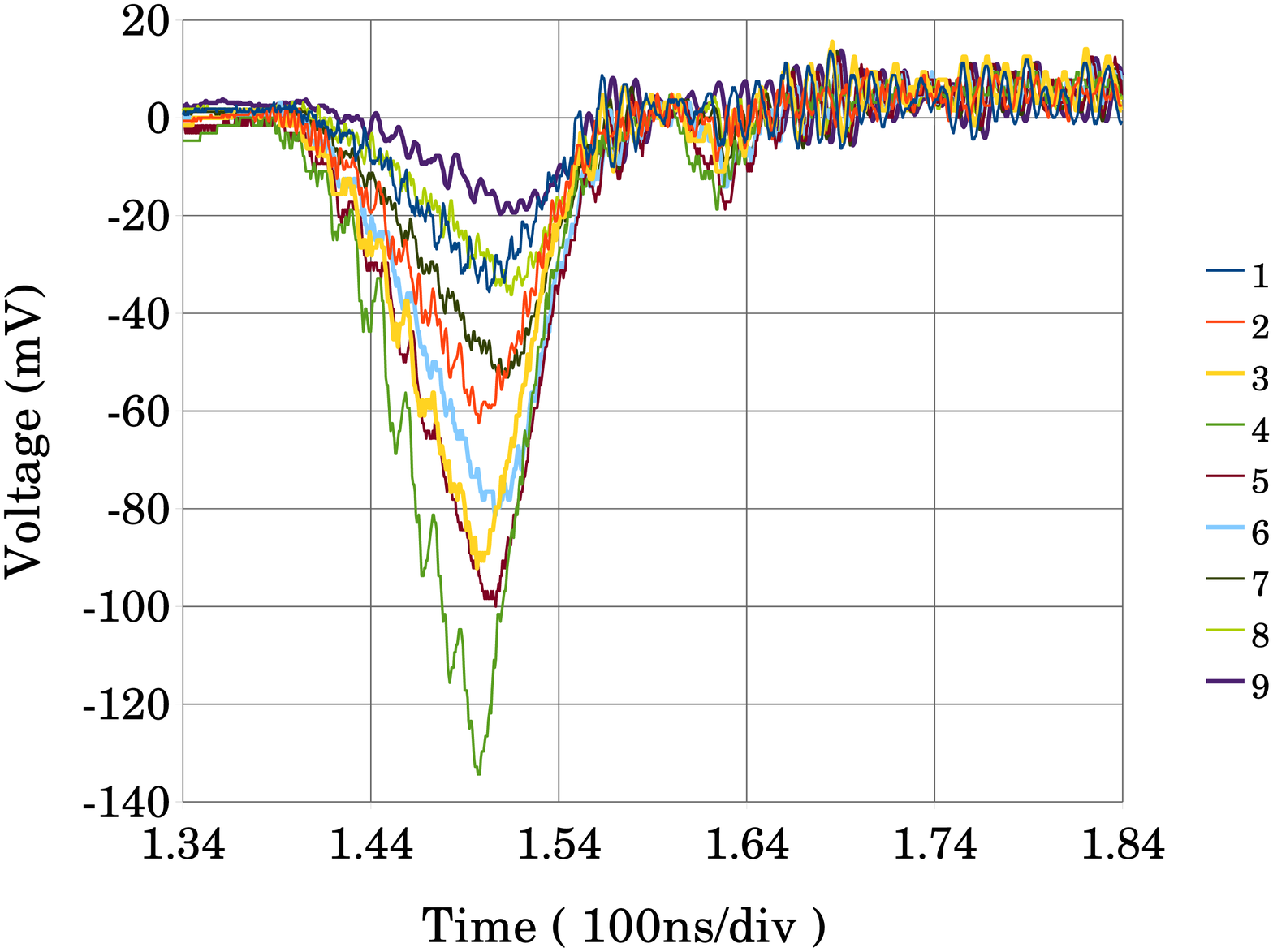}\\
   \includegraphics*[width=78mm]{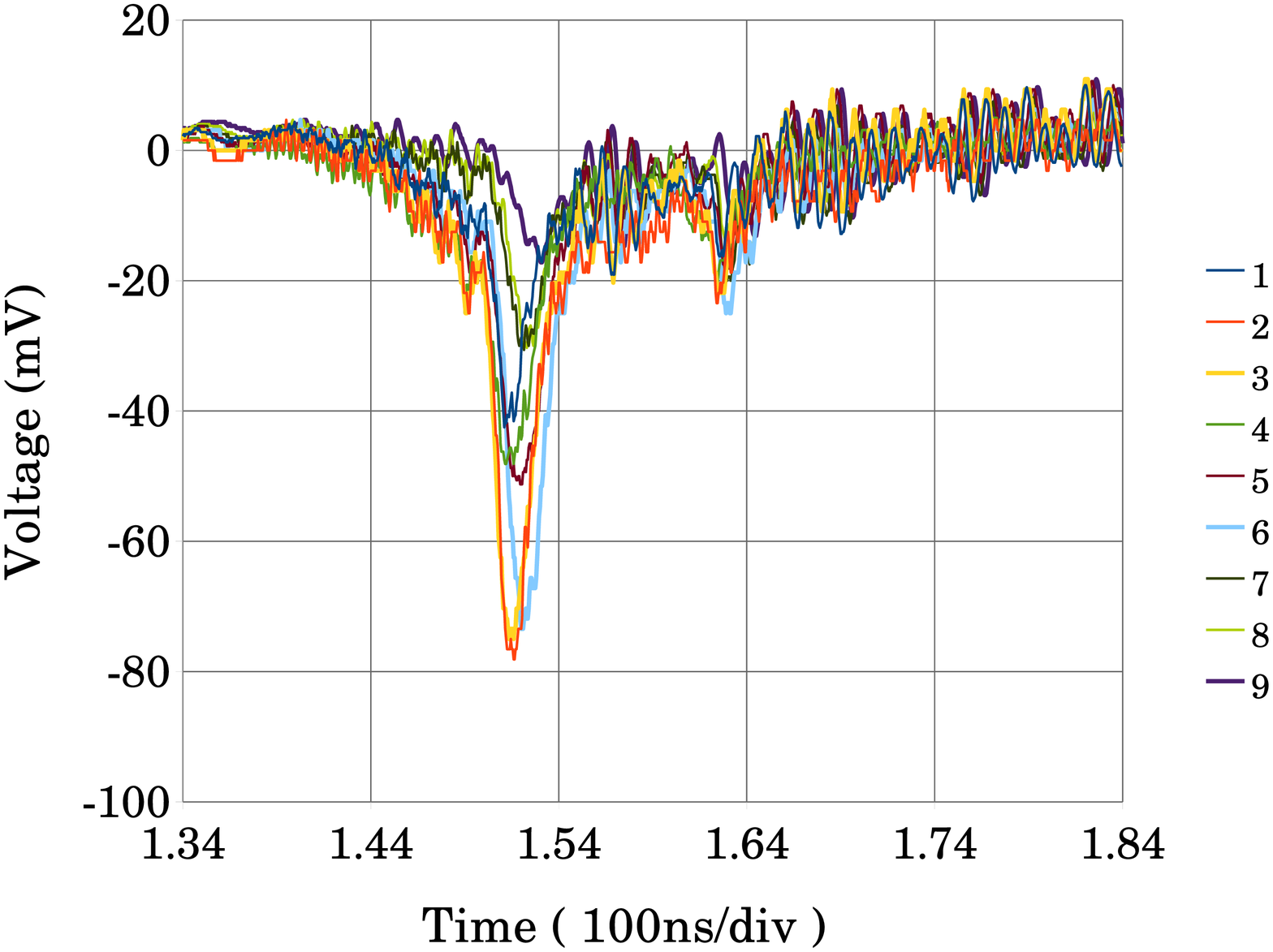}
\vspace{-2.1mm}
   \caption{ Signals from the nine TR-RFA collectors in the aluminum vacuum chamber without grooves.
A train of ten bunches of 5.3~GeV positrons of population \mbox{$1.3{\times}10^{11}$} is followed by one witness bunch  
112~ns after the train. The upper plot shows the case for 
no dipole field. The lower plot shows the effect of a 45~G field, which results in a central depletion zone in the cloud.
}
   \vspace{-2.1mm}
   \label{fig:tr_rfa_signals}
\end{figure}

Two prototype TR-RFAs were installed in the chambers without grooves in 2012.
A bias of +50~V was applied both to the retarding grid and the collectors during this test run.
Initial results can be seen in Fig.~\ref{fig:tr_rfa_signals}, 
where the signal from the TR-RFA in the bare aluminum chamber is shown with and without 
a dipole field of 790~G. The beam in the storage ring consisted of a 10 bunch train of positrons at 5.3~GeV with 14~ns spacing. A witness bunch was also included at a 
delay of 
112~ns following the train. Without magnetic field, collector~4 (near the horizontal center of the beam-pipe) exhibits the largest signal. When the field is turned on, 
all of the signals are reduced in amplitude, but collector~4 is reduced more than the others. This suppression of the central signal with magnetic field has also been 
seen in measurements with standard, time-integrating RFAs~\cite{ref:ipac10jrc}. Such a 
vertical central depletion zone may be due to the effect of beam bunch
kicks increasing the energies of the cloud electrons in the vertical plane of the beam to values 
exceeding the maximum of the true secondary yield curve~\cite{ref:zim1997,ref:jiminez2003}. 
Thus model comparisons may provide sensitivity to the yield curve. Development of detector acceptance
functions in magnetic fields will be important in this modeling project.

\section{Summary}

Time-resolved measurements of electron cloud buildup at {\cesrta} with good time resolution have shown remarkable 
discriminating power for the 
contributing physical processes, distinguishing photoelectron generation characteristics from those of secondary electron 
emission, as well as 
individually identifying the various types of secondary emission. The sensitivity to the kinetic energy 
distribution in the cloud constrains
the production energy distributions of both photoelectrons and secondary electrons. The witness bunch method has 
provided detailed information on
the {\em in situ} vacuum chamber comparisons, including beam conditioning information distinguishing changes in 
photoelectron emission from those of secondary emission.

This summer we installed unconditioned uncoated and TiN-coated aluminum vacuum chambers and recorded witness bunch 
data to be used in determining
their early conditioning characteristics. Following the summer/fall operation as a high-intensity X-ray source, CESR 
will have a dedicated {\cesrta} data-taking
period in November and December, allowing measurement of the uncoated and TiN-coated chambers. 
Four time-resolved retarding field analyzers have been installed in weak dipole magnets with uncoated and TiN-coated aluminum
chambers both grooved and smooth. These will provide the first time-resolved measurements of cloud buildup in magnetic 
fields with fine transverse segmentation
and variable cloud electron energy acceptance thresholds.

\end{document}